\documentclass[sigconf]{acmart}

\copyrightyear{2026}
\acmYear{2026}
\setcopyright{cc}
\setcctype{by}
\acmConference[ICSE-NIER '26]{2026 IEEE/ACM 48th International Conference on Software Engineering}{April 12--18, 2026}{Rio de Janeiro, Brazil}
\acmBooktitle{2026 IEEE/ACM 48th International Conference on Software Engineering (ICSE-NIER '26), April 12--18, 2026, Rio de Janeiro, Brazil}
\acmPrice{}
\acmDOI{10.1145/3786582.3786844}
\acmISBN{979-8-4007-2425-1/2026/04}

\usepackage[linesnumbered,ruled,vlined]{algorithm2e}
\usepackage{textcomp}
\usepackage{xcolor}
\usepackage{subcaption}
\usepackage{multirow}
\usepackage{makecell}
\usepackage{enumitem}
\usepackage{listings}

\begin{document}

\newcommand{\todo}[1]{\textbf{\color{red}{Ravishka: #1} }}

\newcommand{\wei}[1]{\textcolor{blue}{Wei: [#1]}}

\newcommand{\ww}[1]{\textcolor{blue}{Weihang: [#1]}}

\newcommand{\td}[1]{\textcolor{blue}{ToDo: [#1]}}

\newcommand{\zijie}[1]{{\color{purple}[Zijie: #1]}}

\newcommand{\zeqing}[1]{{\textcolor{blue}[Zeqing: #1]}}

\newcommand{\jiajun}[1]{{\color{cyan}[Jiang: #1]}}

\newcommand{\distance}{4pt}
\setlength{\textfloatsep}{\distance}

\newcommand\mycommfont[1]{\small\ttfamily\textcolor{violet}{#1}}
\SetCommentSty{mycommfont}

\lstdefinestyle{Cpp}{ 
	language=C++,
	basicstyle=\scriptsize\ttfamily, 
	breakatwhitespace=false, 
	breaklines=true, 
	captionpos=b, 
	commentstyle=\color[rgb]{0.0, 0.5, 0.69},
	deletekeywords={}, 
	escapeinside={<@}{@>},
	firstnumber=1, 
	frame=lines, 
	frameround=tttt, 
	keywordstyle={[1]\color{blue!90!black}},
	keywordstyle={[3]\color{red!80!orange}},
	morekeywords={String,int}, 
	numbers=left, 
	numbersep=-8pt, 
	numberstyle=\tiny\color[rgb]{0.1,0.1,0.1}, 
	rulecolor=\color{black}, 
	showstringspaces=false, 
	showtabs=false, 
	stepnumber=1, 
	stringstyle=\color[rgb]{0.58,0,0.82},
	tabsize=2, 
	backgroundcolor=\color{white}
}

\title{On-the-Fly Input Adaptation for Reliable Code Intelligence}

\author{Ravishka Rathnasuriya}
\orcid{0009-0005-6129-2865}
\affiliation{%
  \institution{The University of Texas at Dallas
}
  \country{USA}
}
\email{ravishka.rathnasuriya@utdallas.edu}

\author{Wei Yang}
\orcid{0000-0002-5338-7347}
\affiliation{%
  \institution{The University of Texas at Dallas}
  \country{USA}
}
\email{wei.yang@utdallas.edu}

\begin{abstract}

Code language models (CLMs) play a central role in software engineering across both generation and classification tasks. However, these models still exhibit notable mispredictions in real-world applications, even when trained on up-to-date data. Existing solutions address this by retraining the model, modifying its architecture, or re-engineering prompts. These approaches incur high computational cost
requiring substantial effort in data labeling, model updates, and redeployment, and often suffer from poor generalization across tasks and tuning instability across models.

This work proposes an alternative strategy based on on-the-fly input adaptation, which improves model behavior without altering its parameters or requiring additional supervision. The method consists of two stages: input validation, which detects inputs likely to cause mispredictions, and input adaptation, which transforms them using syntax- and semantics-preserving operations to better align with the model’s learned behavior. This dual strategy reduces mispredictions across diverse code understanding tasks, boosting model performance without necessitating retraining. As a scalable and resource‑efficient solution, this framework holds significant promise for high‑stakes applications in software engineering where reliability is critical.

\end{abstract}
\maketitle

\section{Introduction}

Code language models (CLMs) have become essential tools in software engineering tasks such as code generation, summarization, defect prediction, and vulnerability detection~\cite{tian2023fly, naturalattack, Zhang2023Challenging, lu2021codexglue, yefet2020adversarial,hu2023codes,li2021estimating,van2020tailoring,peng2018t,roziere2023code,guo2024deepseek,jiang2024self,evalplus,evalperf}. Trained on large corpora of source code, these models leverage statistical patterns to produce high-quality predictions across a range of inputs. However, despite strong performance on benchmark datasets, CLMs frequently produce incorrect outputs in real-world settings—even when the inputs are syntactically valid and semantically meaningful. Such mispredictions undermine the reliability of CLMs in deployment environments.


A fundamental limitation of CLMs lies in their reliance on surface-level statistical correlations rather than functional understanding. These models treat code as token sequences and do not reason over semantics or program behavior. As a result, minor syntax-preserving changes, such as variable renaming, control flow restructuring, or reordering statements, can lead to inconsistent predictions, despite preserving the underlying program logic~\cite{tian2023fly, naturalattack, Zhang2023Challenging}. This brittleness reflects a deeper misalignment between the model’s learned representations and the structural properties of software.


Existing approaches for addressing such errors typically involve retraining, architectural modifications, or prompt engineering~\cite{yuDataAugmentationProgram2022,xiao2021selfchecking,xiao2022repairing,li2024doce,peng2025perfcodegen,ni2023lever,tian2025fixing}. Retraining requires substantial labeled data, high computational cost, and repeated model updates and redeployment. It also risks overfitting to recent data and is not scalable across evolving tasks or codebases. Architectural changes are model-specific, often incompatible with existing checkpoints, and necessitate full retraining. Prompt tuning, though lightweight, is highly unstable because the performance varies significantly with minor prompt changes and often lacks generalization across tasks and models.


This work proposes a lightweight alternative based on input refinement at inference time. Instead of modifying the model, we adapt the input to better align with the model’s learned decision boundaries. The framework consists of two key components: input validation and input adaptation with integrated search. The \textit{input validation} phase identifies inputs likely to cause model errors. Tailored to different model architectures, this phase includes sub-modules for validating classification models and decoder-based generative models, leveraging uncertainty-based signals appropriate to each. The second phase, \textit{input adaptation}, modifies the flagged inputs using domain-specific transformations to align them with the model’s handling capabilities. This includes discrete code transformations and latent-space perturbations. An integrated search strategy explores multiple adapted variants and selects the one that yields the most reliable output. The overall process reduces mispredictions without modifying model parameters or requiring retraining.



We hypothesize that dynamically adapting inputs at inference time rather than modifying model parameters offers a scalable and generalizable approach to improving the reliability of code language models in practical software engineering settings. By leveraging both structure-preserving transformations and representation-level modifications, this method provides a systematic and model-agnostic mechanism to reduce prediction errors without the computational and operational overhead of retraining.

\section{Research Overview}
\label{problem}

\begin{figure}[!htbp]
\centering
\includegraphics[width=\columnwidth]{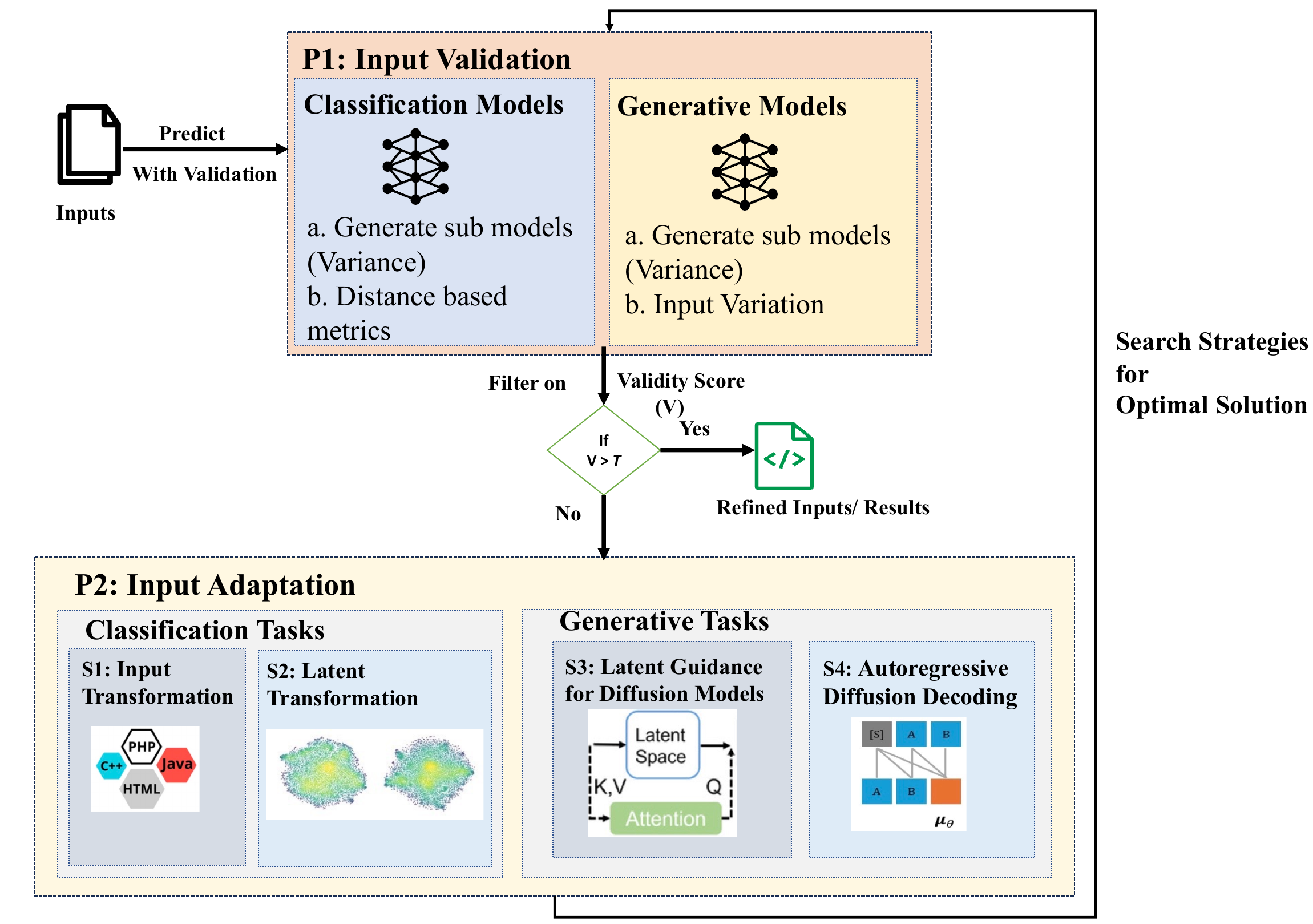}
\caption{Overview of the Proposed Input Adaptation Framework} 
\label{fig:Overview}
\end{figure}

Figure~\ref{fig:Overview} presents an overview of the proposed framework. The framework operates in two phases: Input validation (\textbf{P1}) and input adaptation (\textbf{P2}), followed by a re-inference step guided by a validity score $V$. If $V$ exceeds a task-dependent threshold $T$, the model’s original prediction is retained; otherwise, the input is adapted and reprocessed. 

The framework is designed to handle two families of task inputs. For classification tasks~\cite{tian2023fly, naturalattack, Zhang2023Challenging, lu2021codexglue, yefet2020adversarial,hu2023codes,li2021estimating,van2020tailoring,peng2018t, rathnasuriya2025codeimprove}, inputs consist of function-level code snippets for tasks such as vulnerability detection or defect classification. For generative tasks~\cite{roziere2023code,guo2024deepseek,jiang2024self,evalplus,evalperf}, inputs include natural language specifications, partial programs, or I/O signatures for tasks like synthesis, completion, or repair. These inputs vary in structure, length, and semantics, and often exhibit style-sensitive failure modes. As such, the framework introduces architecture-specific reliability estimation mechanisms that can be applied post-inference.

\label{design}

\textbf{P1: Input Validation.} The objective of the validation phase is to estimate the correctness of the model’s output. This enables post-hoc decisions about whether to accept the prediction or trigger adaptation. Validation is tailored to the model architecture and leverages signals that have been shown to correlate with uncertainty, error likelihood, and distributional fragility.

For \textbf{classification models}, we propose two complementary mechanisms. First, variance-based validation via sub-models generate a small set of perturbed sub-models and measuring output variance provides an efficient proxy for epistemic uncertainty. This draws from ensemble theory and offers a scalable method to estimate prediction stability, particularly in boundary regions of the feature space. Second, representation distance metrics measure the distance between an input’s latent representation and known class prototypes (e.g., via feature-density estimation) can flag examples that are unsupported or atypical. This approach is well-grounded selective classification, and serves as a principled signal of misalignment with the model’s internal class geometry. These mechanisms jointly allow the framework to assign a post-inference validity score without requiring additional supervision or retraining.

For \textbf{generative models}, the validation phase focuses on detecting instability in the decoding process and identifying inputs that may produce invalid or fragile generations. Decoding trajectory analysis uses early-stage decoding signals such as log-likelihood drop-off, or token-level dispersion which can be extracted using auxiliary heads or adapters. These signals are known to correlate with semantic drift, or poor executability. We envision leveraging these low-cost indicators to detect unstable generations before full decoding is completed. Prompt variation consistency applies small perturbations to the input prompt (e.g., reordering elements, rephrasing constraints) can reveal robustness under minor semantic shifts. We propose using self-consistency checks across perturbed versions to measure alignment and stability.

\textbf{P2: Input Adaptation. }The \textbf{P2} addresses inputs
flagged as out-of-scope during validation by applying targeted
transformations to align these inputs more closely with the
model’s handling capabilities. The central hypothesis behind input adaptation is that many model failures stem not from model incapacity, but from mismatches between the input structure and the model’s inductive biases. Rather than retrain the model or tune it for each input variation, we envision a mechanism that \textit{adapts the input to fit the model}, through a series of meaning-preserving transformations that improve model alignment. Adaptation consists of multiple modules from \textbf{S1} through \textbf{S4}, each targeting a distinct failure modality which operate in both classification and generative settings.

\textbf{S1: Input-Space Transformations.} The first adaptation mechanism operates directly in the input space, applying transformations that preserve task semantics while modifying surface features that may mislead the model. These transformations are designed to reduce spurious variance, disambiguate intent, and steer inputs toward regions of higher model confidence without altering their functional meaning. For generative tasks, we propose two transformation classes: (1) Semantic-preserving code transformations, such as variable renaming, control-flow restructuring, and code simplification. These modifications retain the behavior of the original program while reshaping its syntax and structure to better align with the model’s inductive biases. This is particularly effective in code generation, where models are often sensitive to token order and naming conventions and (2) Natural language transformations, including synonym replacement, prompt rephrasing, and constraint reordering. These techniques clarify instruction phrasing and reduce ambiguity, helping the model to focus on intended semantics rather than spurious linguistic cues. Since generative models often exhibit high sensitivity to surface form, such transformations are critical for improving output stability. Across both modalities, the objective of input-space transformation is to preserve meaning while improving the compatibility of the input with the model’s learned decision boundaries. This serves as a lightweight and model-agnostic first step in the adaptation pipeline.

\textbf{S2: Latent-Space Transformations.} While input-space transformations affect the token-level structure, they may not resolve cases where the encoded representation remains misaligned with reliable regions of the model’s internal space. To address this, we propose latent-space perturbation, which adjusts the input embedding prior to decoding. This transformation is guided by the $V$ and follows the principle of confidence-gradient ascent where nudging the latent vector toward a region associated with higher predicted reliability. Importantly, this is achieved under bounded constraints to preserve the semantics of the original input.

The motivation is geometric where models organize input embeddings in a high-dimensional space where certain regions correlate with high prediction confidence. Latent-space adaptation shifts inputs from low-density, unstable regions to those where the model’s decision boundary is more robust. This strategy is particularly effective when surface-level rewrites are insufficient, such as in cases where structurally similar inputs are mapped to distinct internal representations due to sparsity in training coverage.

\textbf{S3: Latent Guidance for Diffusion-Style Generation.} In generative settings that support latent diffusion, we propose a module that conditions the generation process using the reliability signal $V$. During the sampling trajectory, the model receives directional cues in latent space that guide it away from regions associated with unstable or invalid completions. This module is motivated by the observation that diffusion-style generators are inherently underdetermined and often sample from multi-modal distributions. By incorporating  constraints from the validation phase, latent guidance reshapes the sampling path to favor semantically coherent, executable outputs without forcing a deterministic response. Such conditioning introduces a reliability-driven prior over the generative space, which improves outcome stability while retaining diversity.

\textbf{S4: Autoregressive Diffusion Decoding.} For autoregressive generation, early token-level errors often cascade, leading to invalid or incoherent sequences. To mitigate this, we propose a diffusion-inspired autoregressive decoding strategy that periodically revises past generations based on feedback from $V$ and internal confidence signals. This technique treats decoding as a denoising process, where partially completed generations are iteratively refined from coarse-to-fine using internal signals about instability. Unlike standard greedy or beam decoding, this approach allows mid-sequence corrections, which are essential when the generation path diverges due to small prompt shifts or ambiguous instructions. The goal is to interrupt and correct error propagation, especially in tasks like code synthesis or patching, where early inaccuracies can make the entire output unusable.

\textbf{Search strategies for input refinement.} Input adaptation inherently involves choosing among multiple candidate transformations. To make this tractable and effective, we propose a series of guided search strategies, where the validation score $V$ serves as the fitness function. The choice of search method depends on the transformation space. \textbf{Evolutionary search} is suitable for complex input spaces by evolving a population of transformed inputs using mutation operators (e.g., prompt perturbation, identifier mutation). Candidates are scored via $V$, and selection prioritizes inputs likely to yield reliable predictions. In tasks with strong structural rules (e.g., code completion with type constraints), \textbf{constrained decoding} narrows the generation space to syntactically valid candidates only. This approach reduces risk by ensuring that only well-formed completions are considered in the adaptation loop.

\section{Contributions}

In summary, this research makes the following contributions:

\begin{itemize}[leftmargin=*,labelsep=0.5em]

\item \textbf{Multi‑Level Validation Framework:} We propose a unified input validation mechanism that combines complementary metrics to identify inputs likely to trigger errors across classification and generative models in software engineering tasks.

    \item \textbf{Semantic‑Preserving Input Adaptation:} We introduce a suite of input‑space and latent‑space transformation techniques tailored to code and natural‑language prompts. 

    \item \textbf{Validation‑Guided Adaptive Search:} We design a search procedure that uses the validation score $V$ as a guiding signal to explore multiple candidate transformations and select the most reliable representation under a given compute budget. 

    \item \textbf{Model‑Agnostic Evaluation Protocol:} We outline a framework for evaluating reliability improvements across heterogeneous tasks and architectures, illustrating how the approach can be integrated into real‑world deployment pipelines.

    \item \textbf{Scalable, Resource‑Efficient Deployment:} By dynamically refining inputs at inference time, the proposed framework offers a practical alternative to frequent retraining, reducing operational costs and improving reliability in agile and large‑scale software engineering environments.

    \item \textbf{Open Science and Community Impact:} The project advances uncertainty‑aware code intelligence by releasing methodology, validation tools, and adaptation scripts in accordance with open science principles, supporting reproducibility and adoption.
\end{itemize}











\section{Preliminary Results}
\label{study}

We investigate whether uncertainty metrics developed for NLP and vision transfer to pretrained code models across classification and generation tasks~\cite{hendrycks2018baseline,steinhardt2016unsupervised,shannon1948mathematical,monarch2021human,gal2016dropout,lakshminarayanan2017simple,rathnasuriya2025codeimprove,rathnasuriya2025framework,rathnasuriya2025fly,vashurin2024benchmarking}. As a first step toward on-the-fly input adaptation, we evaluate (i) how well these signals detect mispredictions and (ii) how our adaptation modules improve prediction quality without retraining.

\subsection{Effectiveness of Input Validation Using Existing Uncertainty Measurements}

We evaluated a broad set of 13 uncertainty metrics for classification and 16 for generative tasks across multiple pre-trained code models ~\cite{hendrycks2018baseline,steinhardt2016unsupervised,shannon1948mathematical,monarch2021human,gal2016dropout,lakshminarayanan2017simple,rathnasuriya2025codeimprove,vashurin2024benchmarking}. Due to space constraints, we report AUC scores for a subset of four representative metrics for each task, using DeepSeek-Coder-7B and CodeLlama-7B.  For classification, we evaluated vanilla confidence, entropy, Monte Carlo dropout (MCD), and ensemble variance on a vulnerability detection task. For code generation, we used the MBPP+~\cite{dong2025codescore} and HumanEval+~\cite{liu2023your} benchmarks and measured perplexity, entropy, monte-carlo semantic entropy (MCSE), and black-box label probability (BBLabProb)~\cite{vashurin2024benchmarking}. Tables~\ref{tab:uncertainty-metrics} and~\ref{tab:vuldet-auc} summarize the AUC values.

Across both tasks and models, the evaluated uncertainty metrics exhibit limited discriminative power, yielding AUCs close to random guessing (0.50–0.66), with minimal variation across architectures. Notably, BBLabProb performs the worst, while even ensemble-based signals fail to outperform simple entropy-based measures. These results suggest that existing uncertainty metrics lack the precision needed to support reliable abstention or selective prediction in code tasks, motivating the need for task-specific validation strategies as proposed in our framework.

\begin{table}[t]
\centering
\caption{ROC-AUC Scores of Existing Uncertainty Methods on Code Generation Benchmarks ($\uparrow$)}
\label{tab:uncertainty-metrics}
\resizebox{\linewidth}{!}{
\begin{tabular}{lcccc}
\toprule
& \multicolumn{2}{c}{DeepSeek-Coder-7B} & \multicolumn{2}{c}{CodeLlama-7B} \\
\cmidrule(lr){2-3}\cmidrule(lr){4-5}
Metric & HumanEval+ & MBPP+ & HumanEval+ & MBPP+ \\
\midrule
Perplexity   & 0.621 & 0.561 & 0.666 & 0.574 \\
Entropy      & 0.615 & 0.567 & 0.628 & 0.575 \\
MCSE         & 0.612 & 0.569 & 0.575 & 0.534 \\
BBLabprob    & 0.505 & 0.466 & 0.497 & 0.535 \\
\bottomrule
\end{tabular}
}
\end{table}

\begin{table}[t]
\centering
\caption{ROC-AUC of Uncertainty Methods on Vulnerability Detection Classifiers ($\uparrow$)}
\label{tab:vuldet-auc}
\begin{tabular}{lcc}
\toprule
Metric & DeepSeek\textendash Coder\textendash 7B & CodeLlama\textendash 7B \\
\midrule
Vanilla  & 0.621 & 0.615 \\
Entropy  & 0.617 & 0.616 \\
MCD      & 0.621 & 0.615 \\
Ensemble & 0.580 & 0.559 \\
\bottomrule
\end{tabular}
\end{table}

\subsection{Effectiveness of Input Adaptation}

We evaluated the impact of our adaptation strategies corresponding to \textbf{S1}  and \textbf{S2} on classification tasks performance using CodeBERT and GraphCodeBERT. For both models, we leveraged a vulnerability detection task~\cite{zhou2019devign} and compared the base model accuracy to that obtained after adaptation. Table~\ref{tab:adaptation-types} summarizes the effectiveness of adaptation strategies. 

Both adaptation strategies substantially improve model performance over the base, with latent-space transformation yielding up to 13.4\% improvement for CodeBERT and 5.3\% for GraphCodeBERT. This confirms that semantics-preserving rewrites (\textbf{S1}) and representation-level adjustments (\textbf{S2}) can shift inputs toward more reliable decision regions, thus validating our hypothesis that input-level intervention, when guided by uncertainty, is a viable alternative to model retraining.

\begin{table}[t]
\centering
\caption{Effectiveness of Input Adaptation for Pretrained Models on Vulnerability Detection ($\uparrow$)}
\label{tab:adaptation-types}
\begin{tabular}{lcc}
\toprule
Adaptation Type & CodeBERT & GraphCodeBERT \\
\midrule
Base                    & 63.36 & 62.99 \\
Input Transformation    & 71.52 & 65.26 \\
Latent Transformation & 76.75 & 68.32 \\
\bottomrule
\end{tabular}
\end{table}

\subsubsection{Semantic Preservation and Runtime Overhead}

A key objective of our framework is to ensure that input transformations preserve the semantic intent and functional behavior of code inputs. Across all experiments, transformations applied in \textbf{S1} and \textbf{S2} consistently preserved original functionality, confirmed by behavior-preserving diffs and output equivalence checks. 

In terms of runtime, the framework is practical for deployment: Input transformations incur per-input adaptation time between 49.92s and 59.4s, depending on model and search strategy. Latent transformations  are significantly faster, with overheads of 2–3 seconds per input, making them well-suited for real-time systems. By avoiding retraining or fine-tuning, the framework offers a resource-efficient and scalable solution for improving model reliability—making it suitable for agile pipelines and large-scale production deployments.

\section{Future Work and Research Plan}
\label{discussion}





Having established the foundational impact of uncertainty metrics on code models across both classification and generation tasks, our future work is organized into a focused short-term execution plan and a broader long-term research vision.

\textbf{Short-Term Plans. } In the short term, our focus is on disseminating early findings and operationalizing the framework proposed in this study. (1) \textit{Dissemination of Findings:} We plan to publish our empirical results on the effectiveness of existing uncertainty metrics in code models to inform the research community of the current limitations and motivate the need for model-agnostic input refinement strategies; (2) \textit{Framework Implementation:} We will complete the full implementation of the proposed uncertainty-driven adaptation pipeline, including both validation metrics (e.g., distance-based, variance-based, decoding stability) and adaptation techniques (e.g., semantic-preserving code transformations, latent-space perturbations, guided decoding); (3) \textit{Task Prioritization:} We will begin with classification tasks where validation and adaptation can be more precisely evaluated using accuracy-risk trade-offs and progressively extend to generative tasks, where metrics such as pass@k and execution correctness provide additional insights; and (4) \textit{Diverse Evaluation:} We also plan to integrate multiple evaluation metrics beyond AUC, including AUROC, AUARC, coverage-risk curves, and misprediction localization, to better quantify the benefits of adaptation under operational constraints.  The objective is to complete this implementation and evaluation cycle within a 12 month window, building a reproducible and extensible platform for adaptation-aware model interaction.

\textbf{Long-Term Plans.}  In the longer term, we aim to broaden the scope of this work to new settings and practical deployments. (1) \textit{Task Expansion:} We will extend our framework to a broader range of classification and generative tasks, including bug repair, type inference, dataflow classification, and constraint-based synthesis; (2) \textit{Black-Box Integration:} We plan to explore the use of our framework in black-box inference settings, where only model APIs are available. In these cases, we will develop confidence proxies and behaviorally grounded signals (e.g., execution traces, token agreement) to guide abstention and adaptation without access to internal logits or representations; (3) \textit{Runtime Integration in Coding Agents:} A major long-term goal is to integrate this framework into real-time code intelligence systems, such as coding assistants, IDE plugins, and automated PR reviewers. This will require optimizing latency, caching intermediate validation results, and creating adaptation strategies that are lightweight and reversible; and (4) \textit{Open-Source Toolkit:} We also plan to release a modular, open-source toolkit that supports plug-and-play adaptation strategies and can be integrated into any existing inference pipeline. Our long-term roadmap is aligned with a 12 to 18 month development cycle, with public releases planned at major milestones. Together, these goals aim to transform the way deployed models interact with inputs shifting the focus from static model outputs to adaptive input processing pipelines that enhance robustness in real-world environments. 

\section{ACKNOWLEDGMENTS}
This work was partially supported by NSF grants NSF CCF2146443 and Amazon Trust AI Research Award.

\clearpage

\bibliographystyle{ACM-Reference-Format}
\bibliography{main}
\end{document}